\documentclass[aps, prb, showpacs, preprint , amssymb, amsmath, article, superscriptaddress, unsortedaddress]{revtex4}
\usepackage{graphicx}
\usepackage{epsfig}
\usepackage{ifpdf}
\usepackage{subfigure}
\usepackage{hyperref}
\include{graphics}

\begin{document}

\title{Strain effect on quantum conductance of graphene nanoribbons from maximally localized Wannier functions }
\author{R. Rasuli}
\affiliation{Sharif University of Technology, Department of Physics,
P.O. Box 11365-9161, Tehran, Iran.}
\author{ H. Rafii-Tabar}
\email{rafii-tabar@nano.ipm.ac.ir}
\affiliation{Computational
Physical Sciences Research Laboratory, Department of Nano-Science,
Institute for Research in Fundamental Sciences (IPM), P.O. Box
19395-5531, Tehran, Iran}
\affiliation{Department of Medical Physics
and Biomedical Engineering and Research Center for Medical
Nanotechnology and Tissue Engineering, Shahid Beheshti University of
Medical Sciences, Evin, Tehran, Iran}

\author{A. Iraji zad}
\affiliation{Sharif University of Technology, Department of Physics,
P.O. Box 11365-9161, Tehran, Iran.}
\affiliation{Institute for
Nanoscience and Nanotechnology (INST), Sharif University of
Technology, P.O. Box 14588-89694, Tehran, Iran}

\date{\today}

\begin{abstract}
Density functional study of strain effects on the electronic band
structure and transport properties of the graphene nanoribbons (GNR)
is presented. We apply a uniaxial strain ($\varepsilon$) in the $x$
(nearest-neighbor) and $y$ (second nearest-neighbor) directions,
related to the deformation of zigzag and armchair edge GNRs (AGNR
and ZGNR), respectively. We calculate the quantum conductance and
band structures of the GNR using the Wannier function in a strain
range from $-8\%$ to $+8\%$ (minus and plus signs show compression
and tensile strain). As strain increases, depending on the AGNR
family type, the electrical conductivity changes from an insulator
to a conductor. This is accompanied by a variation in the electron
and hole effective masses. The compression $\varepsilon_x$ in ZGNR
shifts some bands to below the Fermi level ($E_f$) and the quantum
conductance does not change, but the tensile $\varepsilon_x$ causes
an increase in the quantum conductance to $10e^2/h$ near the $E_f$.
For transverse direction, it is very sensitive to strain and the
tensile $\varepsilon_y$ causes an increase in the conductance while
the compressive $\varepsilon_y$ decreases the conductance at first
but increases later.
\end{abstract}
\newpage
\maketitle
\section{Introduction}\label{Introduction}
The discovery of graphene in 2004 and its potential application in
nanotechnology has attracted considerable interest
\cite{interest1,interest2,expstrain1}. The unique properties of
graphene due to massless Dirac fermions, high surface to volume
ratio, high crystal quality in 2D, and high mechanical strength
recommend it as a promising material for technologies such as
nanoelectromechanical systems (NEMS) and nanosensors
\cite{application,newgr, diracfermion, halleff}. Use of graphene in
nanodevices is accompanied by induced strains and stresses. These
may originate from phonon-induced lattice vibrations, lattice
mismatched film growth, or applied external stress, and it is
therefore necessary to take them into account.

Recently, research has focused on strained graphene in order to
study the effect of strain on the electronic structure properties
\cite{strain1,strain2,strain3,rafiitabar2}. Deformations in a
graphene sheet affect its electron transport by decreasing its
conductance at low densities, making electrons behave in the same
way as in a nanoribbon or a quantum dot \cite{strain3}. Theoretical
studies of the strain effects, followed by recent experiments, have
attempted to understand and control the interplay between the strain
and the electrical transport in graphene \cite{expstrain1,
expstrain2}.

In a nanoribbon, the wave function is localized in two dimensions.
Because of the finite size of a GNR, Bloch orbitals cannot be used
directly in electronic transport calculations, and quantum
conductance should be computed via lattice Green's function and
localized orbital representation of the electronic states in real
space. Here, we employ maximally localized Wannier functions (MLWF)
to compute the electrical conduction properties
\cite{souza2002mlw}.

In this paper we perform an $ab$ $initio$ study of the strain effect
on a flat GNR as the structural motif for carbon nanotubes, and
other graphene-based materials \cite{saitobook}. We calculate the
coherent transport properties of nanostructures from first
principles by combining them with the Landauer approach \cite{qc}.
The connection between first principles and Landauer approach is
provided by using MLWF representation. Using MLWFs, we calculate the
band structure  and the quantum conductance of a GNR under various
strains. The results show that the transport properties of a GNR
depend on the strain. As the strain increases, depending on the AGNR
family type, the electrical conductivity changes from an insulator
to a conductor, and this is accompanied by a variation in the
electron and hole effective masses. The compression $\varepsilon_x$
in the ZGNR shifts some bands to below the $E_f$ and the quantum
conductance does not change, but the tensile $\varepsilon_x$ causes
an increase in the quantum conductance to $10e^2/h$ near the $E_f$.
For transverse direction, the ZGNR is very sensitive to the strain,
and the tensile $\varepsilon_y$ causes an increase in the
conductance while the compressive $\varepsilon_y$ decreases it at
first, but increases later.


\section{computational details}
\label{experiment} 
\begin{figure*}[htp]
\begin{center}
  \subfigure[  ]{\label{arm}\includegraphics[angle=0,scale=0.25]{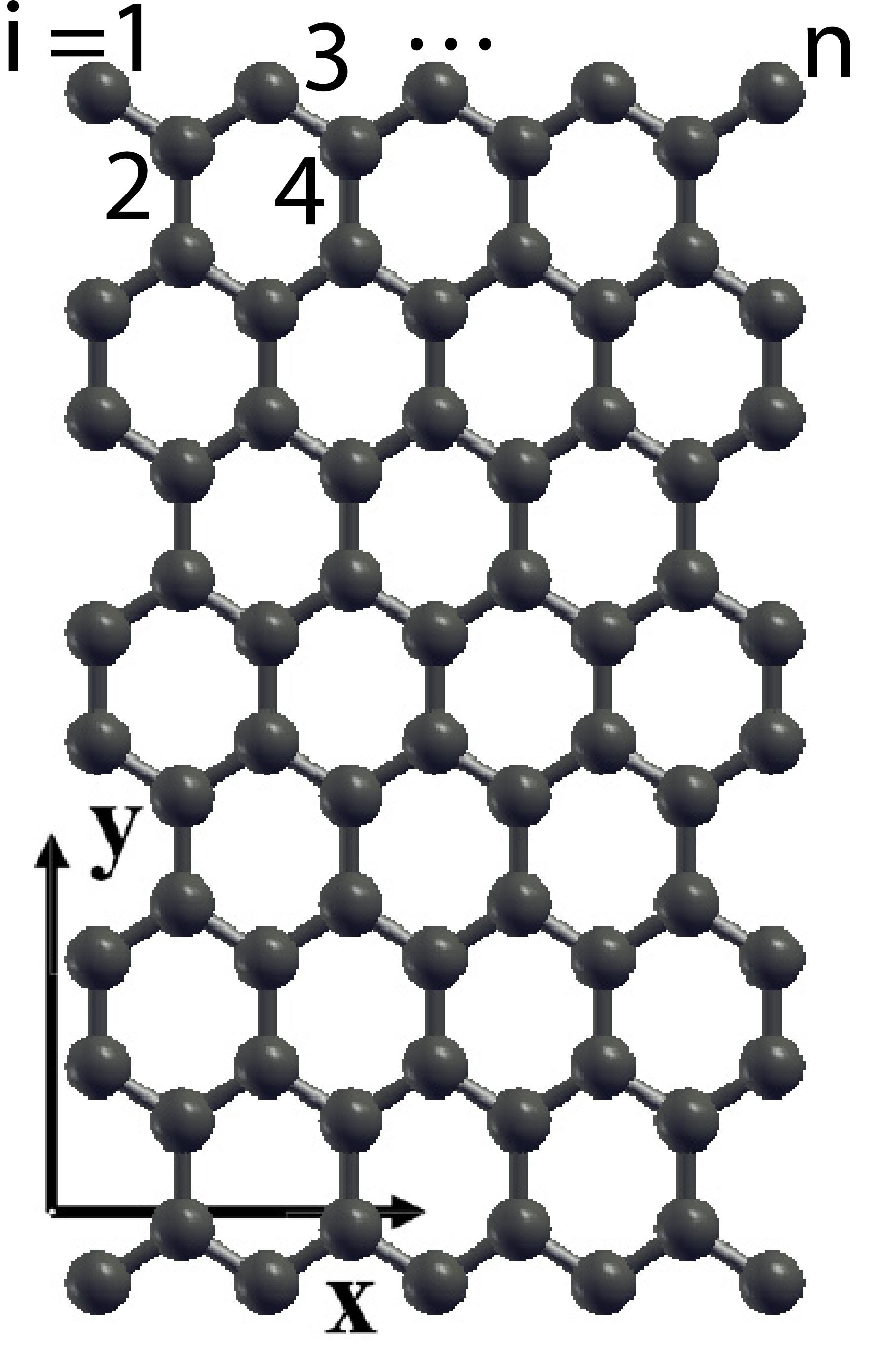}}
  \subfigure[  ]{\label{zig}\includegraphics[angle=0,scale=0.25]{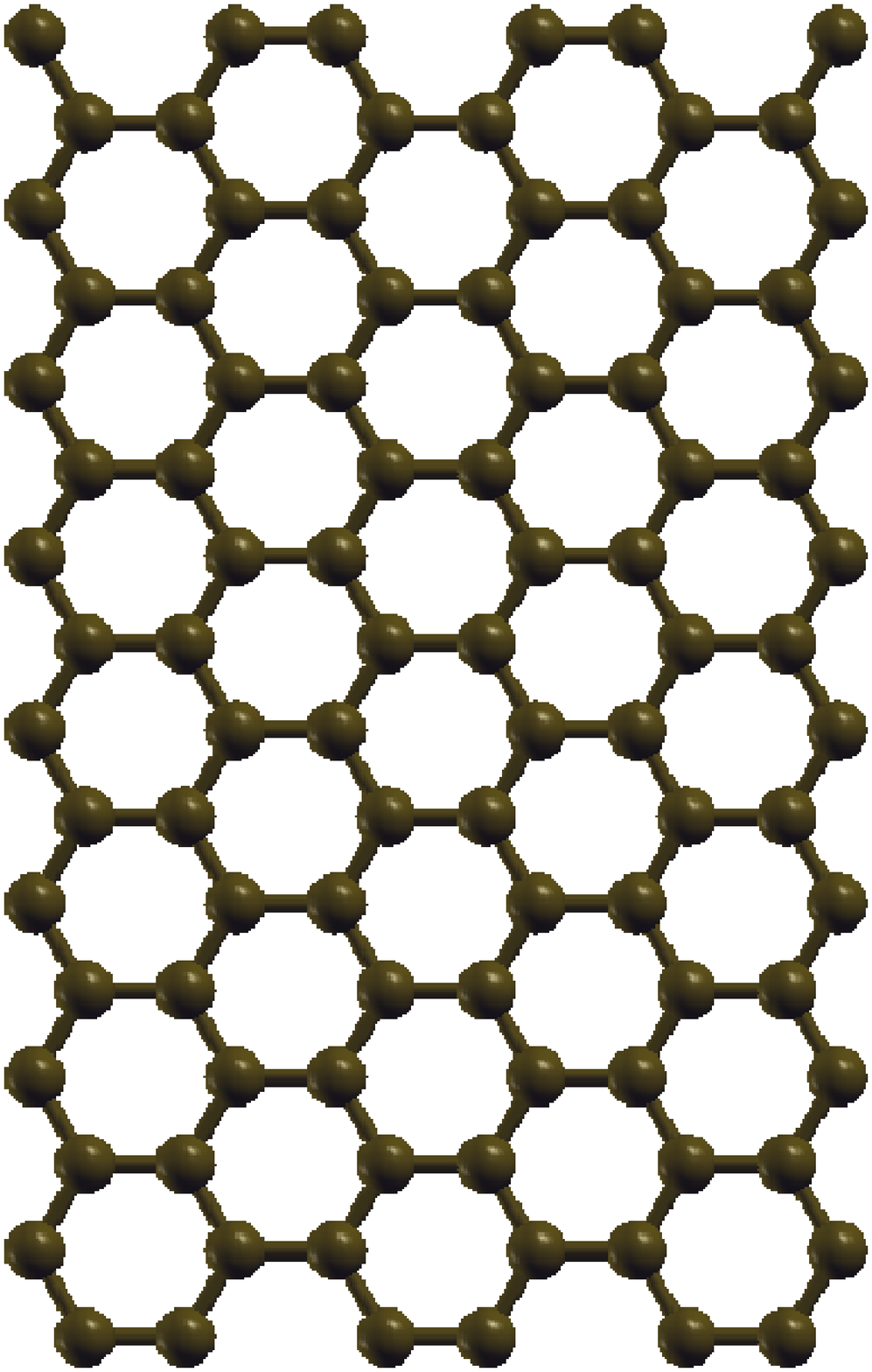}}
\end{center}
\caption{\label{supercell} Graphene nanoribbon supercell.}
\end{figure*}

First-principles DFT-based calculations, which
have been successfully applied to the study of graphene
\cite{sucdft}, have been performed in this paper. All our DFT
calculations were carried out with the PWSCF code \cite{pw, pwnew},
within the local density approximation (LDA) and using the
ultrasoft Rabe-Rappe-Kaxiras-Joannopoulos pseudopotentials
\cite{ultsoft}. We use an MLWF basis set which can be obtained by a
unitary transformation of the occupied ground-state plane wave
eigenfunctions of the density functional theory with the cutoff
energy of 340 eV. The total system consists of 12 and 18 C atoms
(graphene supercell) in a slab geometry in two dimensions (Fig.
\ref{supercell}) and with a distance of 14 {\AA} between the
adjacent layers. The sampling of the Brillouin zone is done using a
$1\times1\times8$ Monkhorst-Pack grid \cite{mgrid}, which is tested
to give converged results for all properties that are calculated.
MLWFs are computed following the method of Marzari and Vanderbilt
(MV) using $wannier90$ code \cite{marzariwan, wan90}. For entangled
energy bands, the method of Souza, Marzari and Vanderbilt (SMV) is
used \cite{souza2002mlw}. The coordinate optimization is done via
conjugate gradient molecular dynamics with geometry constrains in
induced strain direction, and the maximum atomic displacement is
limited to 0.1 {\AA} for each molecular dynamics step in order to preserve
the stability of calculations.

\section{results and discussion }
\label{AGNR} The structure studied in this work is made from a
graphene sheet that is sheared along a zigzag or an armchair edges
and passivated by hydrogen atoms (Fig. \ref{supercell}). Recently,
the effects of edge and size on the GNR band structures and also the
effect of charge transfer from substrate to a graphene layer have
been studied \cite{sbst,ribbonedge}. Here, we have neglected the
charge transfer from substrate, and the edge and size effects, and
have only considered the strain effects. We represent the
deformation of the GNR by the strain ($\varepsilon_i$) defined as
$\Delta a_i/a_i$ where $a_i$ is the strain induced superlattice
constant in the $i=x,y$ directions. To study the strain effects, we
consider the longitudinal and transverse strains separately i.e., in
the $x$ and $y$ directions.

\begin{figure}
\center
\includegraphics[angle=0,scale=0.25]{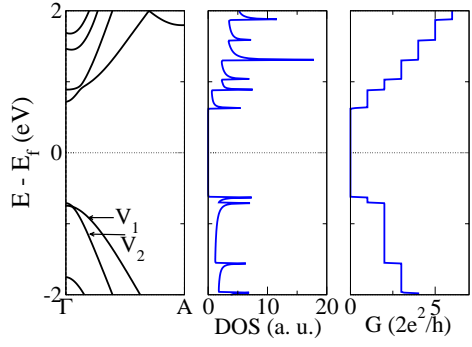}
\caption{\label{armrelax} Band structures, density of states and
quantum conductance of a 9-AGNR at zero strain.}
\end{figure}

First we calculate the electronic structure of the AGNR at zero
strain. An AGNR with widths $w$ = 9, 10, and 11 is chosen to
represent three typical families corresponding to 3$n$, 3$n$+1, and
3$n$+2 (Fig. \ref{supercell}), respectively, similar to the previous
works \cite{3family1, 3family2}. We present every family with $w$
key as $w$-AGNR. Fig. \ref{armrelax} presents the band structures,
the density of states and the quantum conductance of a 9-AGNR.
Neglecting the edge effects, due to passivation, $Van$ $der$ $Waals$
bonds, that create band dispersion in out of the basal plane, are
responsible for band mixing. The band structure is spanned by the
bonding combination of $sp^2$ orbitals (the graphitic backbone) and
$p_z$ orbitals ($\pi$ and $\pi^*$ bands). The hopping integral of
the $\pi$ bands dispersion near the $E_f$ determines the band gap,
the transport property, and the effective mass of the charge carrier. As shown in
Fig. \ref{armrelax}, at $\Gamma$ point, there is a direct gap of 1.4
eV between the valence and conduction bands. In the valence band,
the heavy hole band ($V_1 $ subband) is separated from the light
hole band ($V_2 $ subband). Van Hove singularity in the density of
states shows this band splitting.

Using retarded and advanced Green's functions, based on the Landauer
formalism, quantum conductance is computed. MLWFs, evaluated by a
combination of plane-wave electronic structure, is applied for
further conductance calculation. To construct the wave functions we
select an energy window with $E \in [-6,6]$ eV around the $E_f$
(taken as the reference zero). This energy window contains all the
occupied and empty states. Fig. \ref{armrelax} shows the quantum
transport of the 9-AGNR and its value near the $E_f$ is zero. Here,
conductance includes sum over all transmission possibilities that
are between 0 and 1. Therefore, there are no channels to connect the
two sides of the 9-AGNR near the $E_f$.

\begin{figure*}[htp]
\begin{center}
  \subfigure[ : Strain in $x$ direction for three families of ribbons]{\label{3bandxarm}\includegraphics[angle=0,scale=0.28]{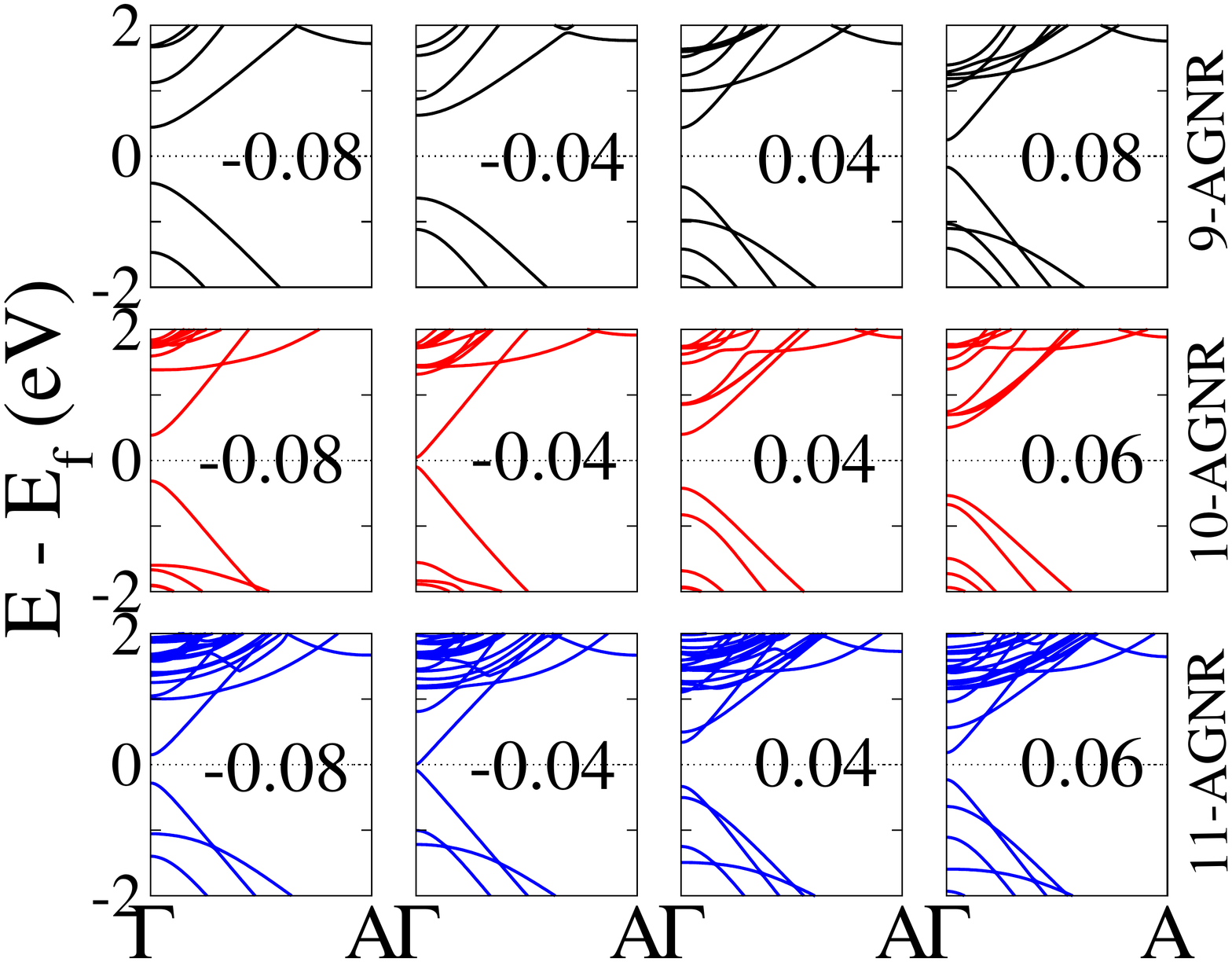}}
  \subfigure[ : Strain in $y$ direction for three families of ribbons]{\label{3bandzarm}\includegraphics[angle=0,scale=0.28]{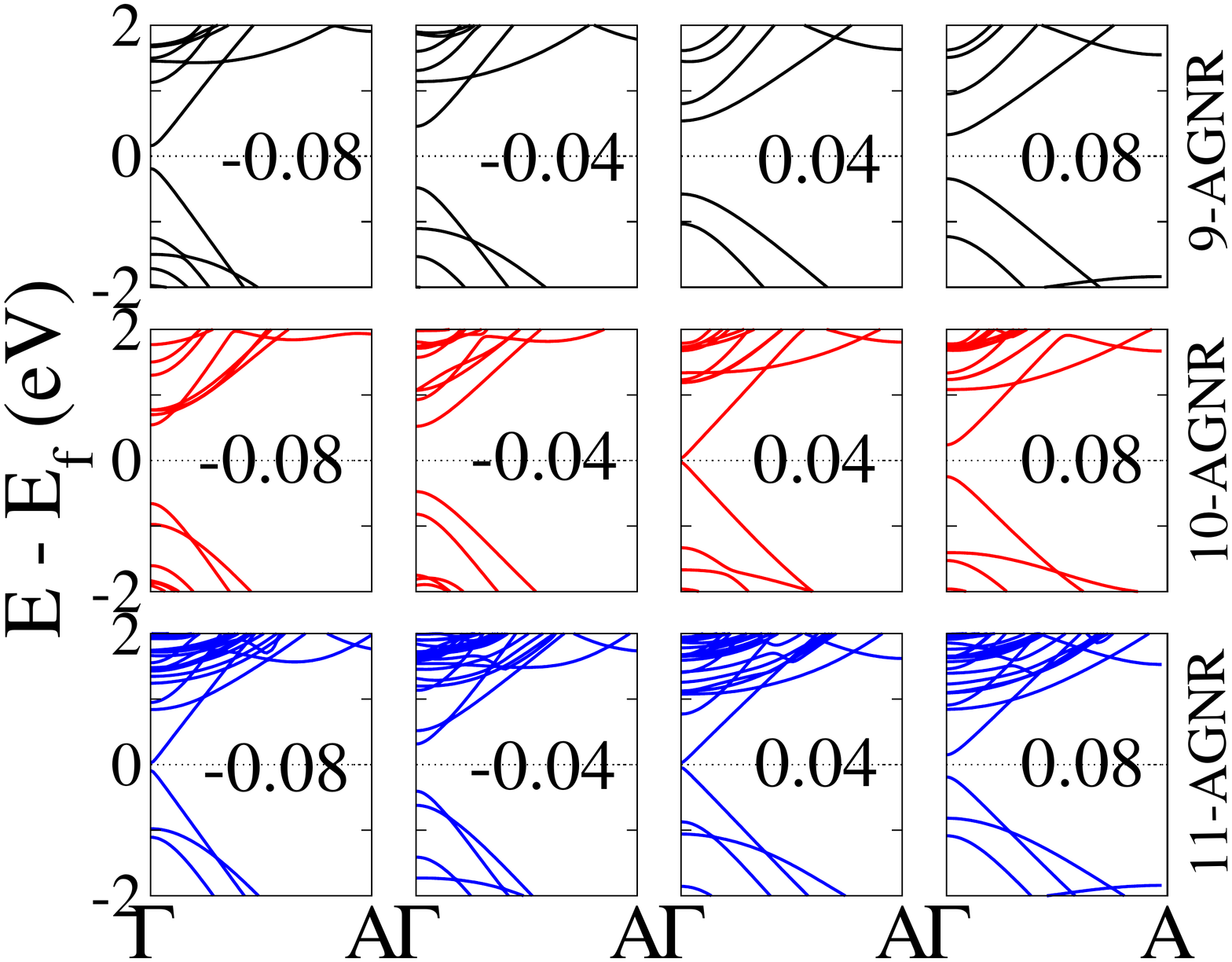}}
\end{center}
\caption{\label{bandarm} Band structures of $n$-AGNR for three
families of ribbon under various strains.}
\end{figure*}

\begin{figure*}[htp]
\begin{center}
  \subfigure[ : W = 9]{\label{9AGNRgap}\includegraphics[angle=0,scale=0.17]{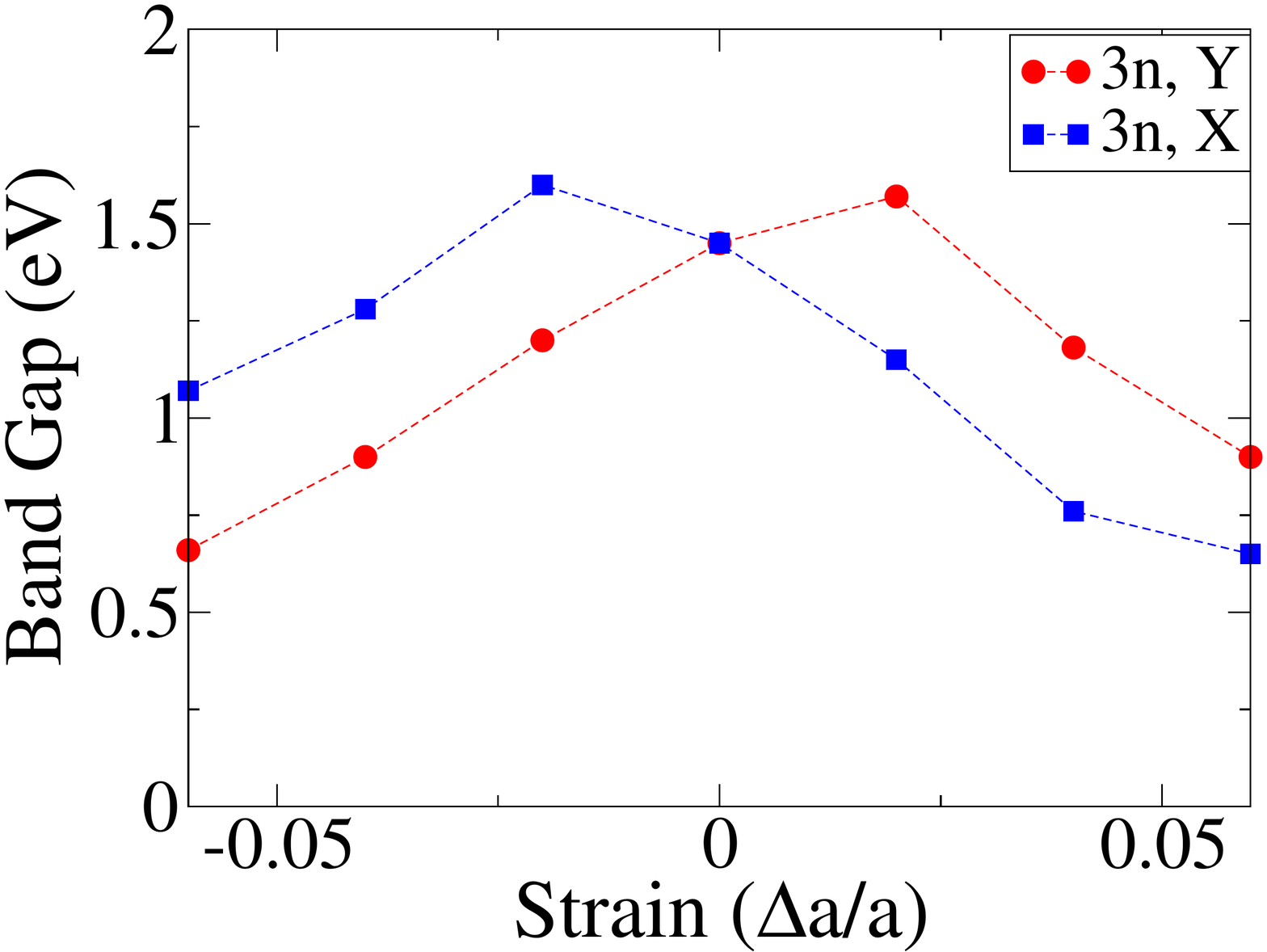}}
  \subfigure[ : W =10]{\label{10AGNRgap}\includegraphics[angle=0,scale=0.17]{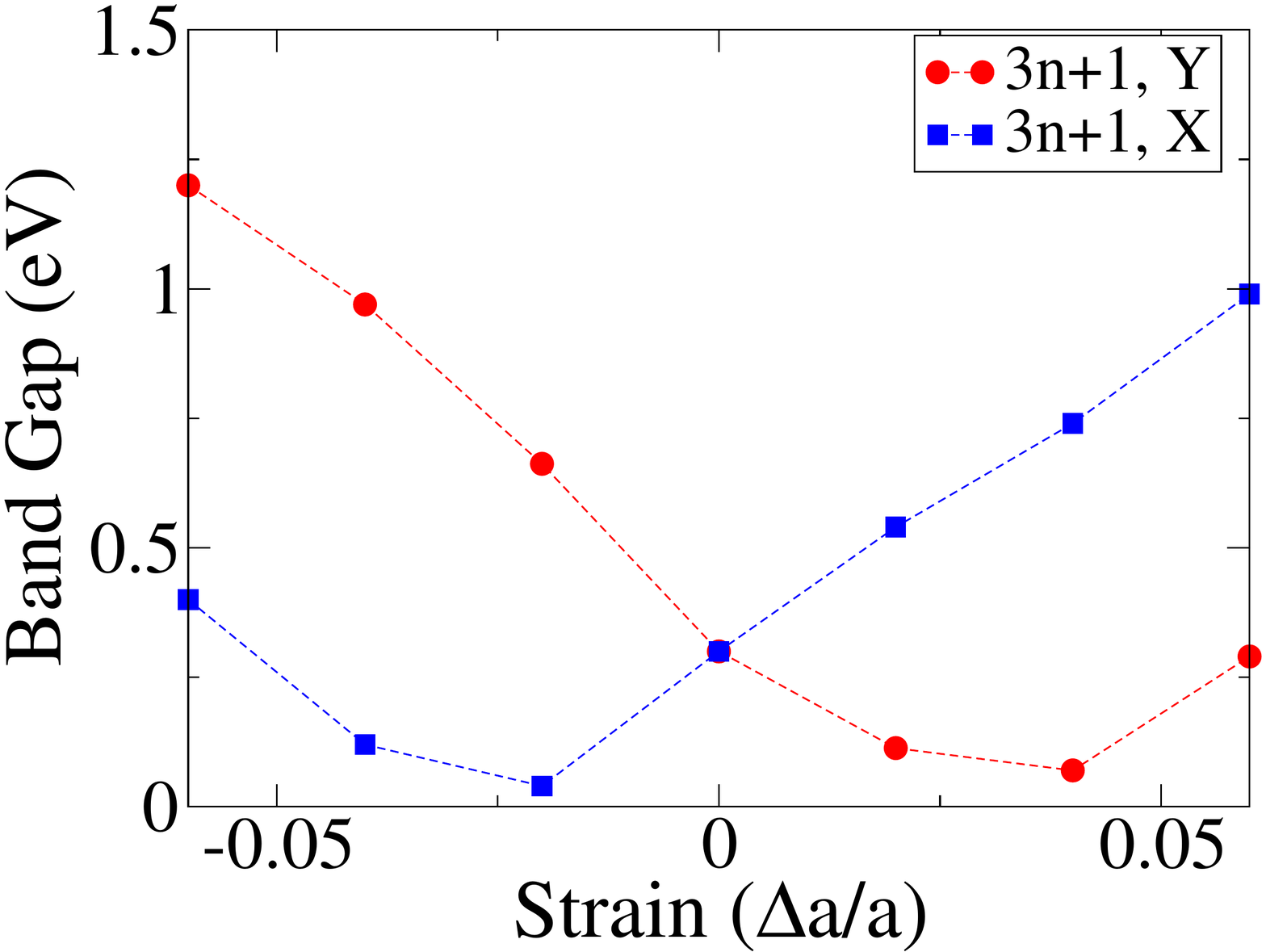}}
  \subfigure[ : W=11 ]{\label{11AGNRgap}\includegraphics[angle=0,scale=0.17]{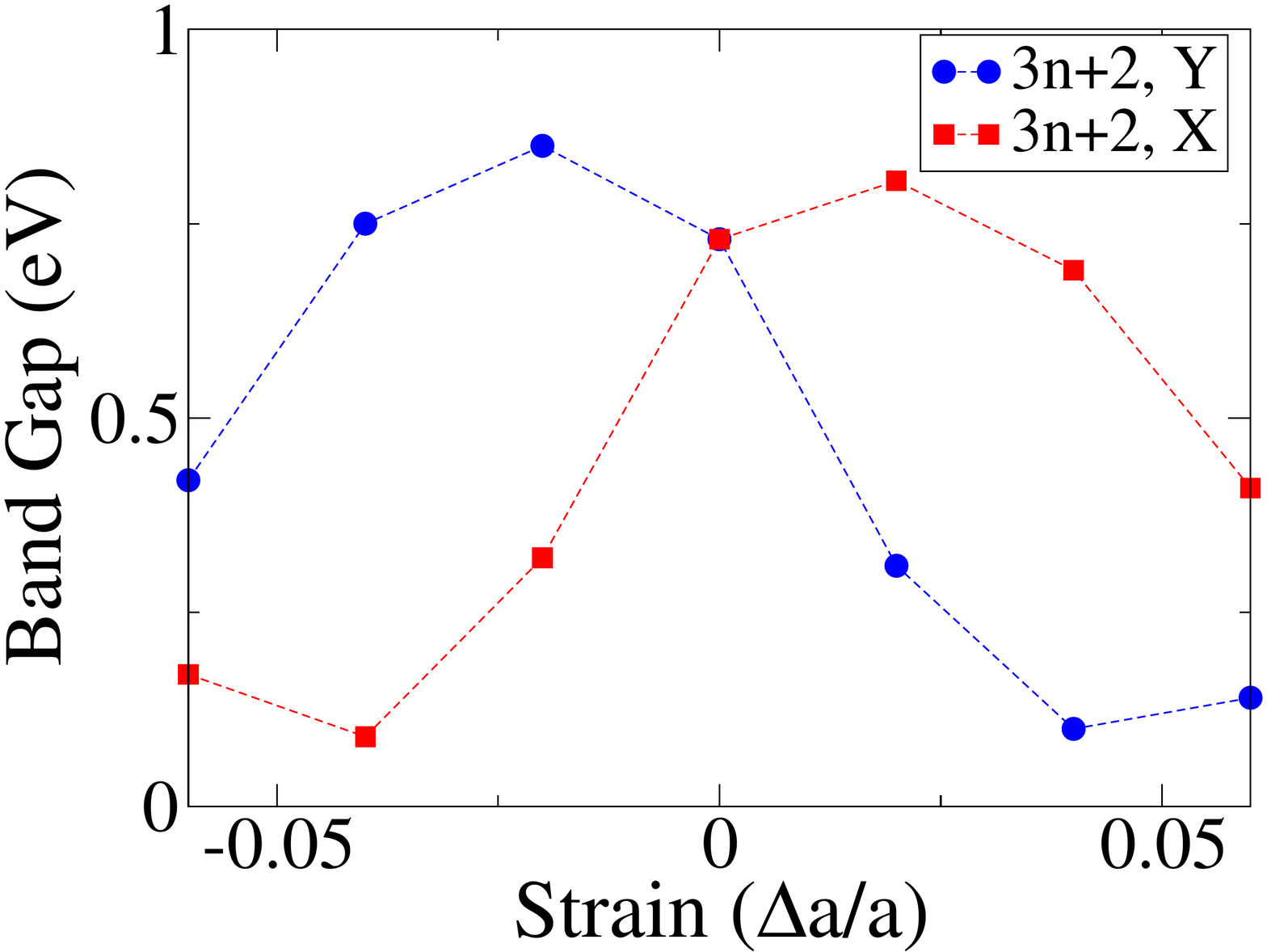}}
\end{center}
\caption{\label{gap} Energy gap of an AGNR under various strains for
$w$ = 3$n$, 3$n$+1 and 3$n$+2.}
\end{figure*}

Fig. \ref{bandarm} shows the band structures of three families of
AGNR under various uniaxial tensile ($\varepsilon>0$) and
compressive ($\varepsilon<0$) strains. The band structures of the
AGNR families are sensitive to the strain, and the band gap changes
non-symmetrically for the tensile and the compressive strains as
shown in Fig.\ref{gap}. For the 3$n$ family in Fig.\ref{9AGNRgap}
the tensile $\varepsilon_y$ and the compressive $\varepsilon_x$
increase the band gap at first and then reach a maximum value at a
strain of about $\varepsilon_x=-0.02$ and $\varepsilon_y=0.02$, and
finally the application of further strain decreases the band gap. Band
gap variation for the 3$n$+1 group is shown in Fig. \ref{10AGNRgap}.
It suppressed to a narrow gap when the induced strain reaches
$\varepsilon_y=0.04$ and $\varepsilon_x=-0.02$, while the compressive
strain in $y$ direction and the tensile strain in $x$ direction
increase the band gap to above 1 eV. Band gap variation for the
3$n$+2 family is completely different, and has a maximum and a
minimum (Fig. \ref{11AGNRgap}). The tensile $\varepsilon_y$
decreases the gap, while the compressive strain causes it to increase
the gap. The corresponding maximum gap occurs at
$\varepsilon_x=0.02$ and $\varepsilon_y=-0.02$ and the minimum is at
$\varepsilon_x=-0.04$ and $\varepsilon_y=0.04$. As shown in Fig.
\ref{gap}, all the behavior in the $y$ direction is inverse of that
in the $x$ direction, and our results demonstrate that the band gap
does not vanish under any condition.

Another result from Fig. \ref{3bandxarm} and \ref{3bandzarm} is the
variation in the effective mass of fermions as the strain induced
curvature of bands at $\Gamma$ point, that is related to the
effective mass, changes. When the tensile strain becomes
compressive, in addition to variation in the effective mass, the
heavy hole band near the band edge ($V_1$ subband) exchanges with
the light hole band ($V_2$ subband). The variation in the effective mass is
different for $w$ =9 and 10, 11. The deformation of the 9-AGNR only causes a
variation in the effective mass, but for the 10 and 11-AGNR family it
can decrease the fermion's effective mass to near zero (Fig.
\ref{bandarm} for $\varepsilon_x=-0.04$ and $\varepsilon_y=0.04$).
In this case, dispersion relation near the $E_f$ is converted to
quasi-linear by a narrow gap. In addition, band deformation, which
depends on strain, results in two types of excitons; first type
relates to heavy electron-hole, and the second corresponds to light
electron-hole.

\begin{figure*}[htp]
\begin{center}
  \subfigure[ : The Landauer quantum conductance variation in the $x$ and $y$ direction.]{\label{9qcAGNR}\includegraphics[angle=0,scale=0.28]{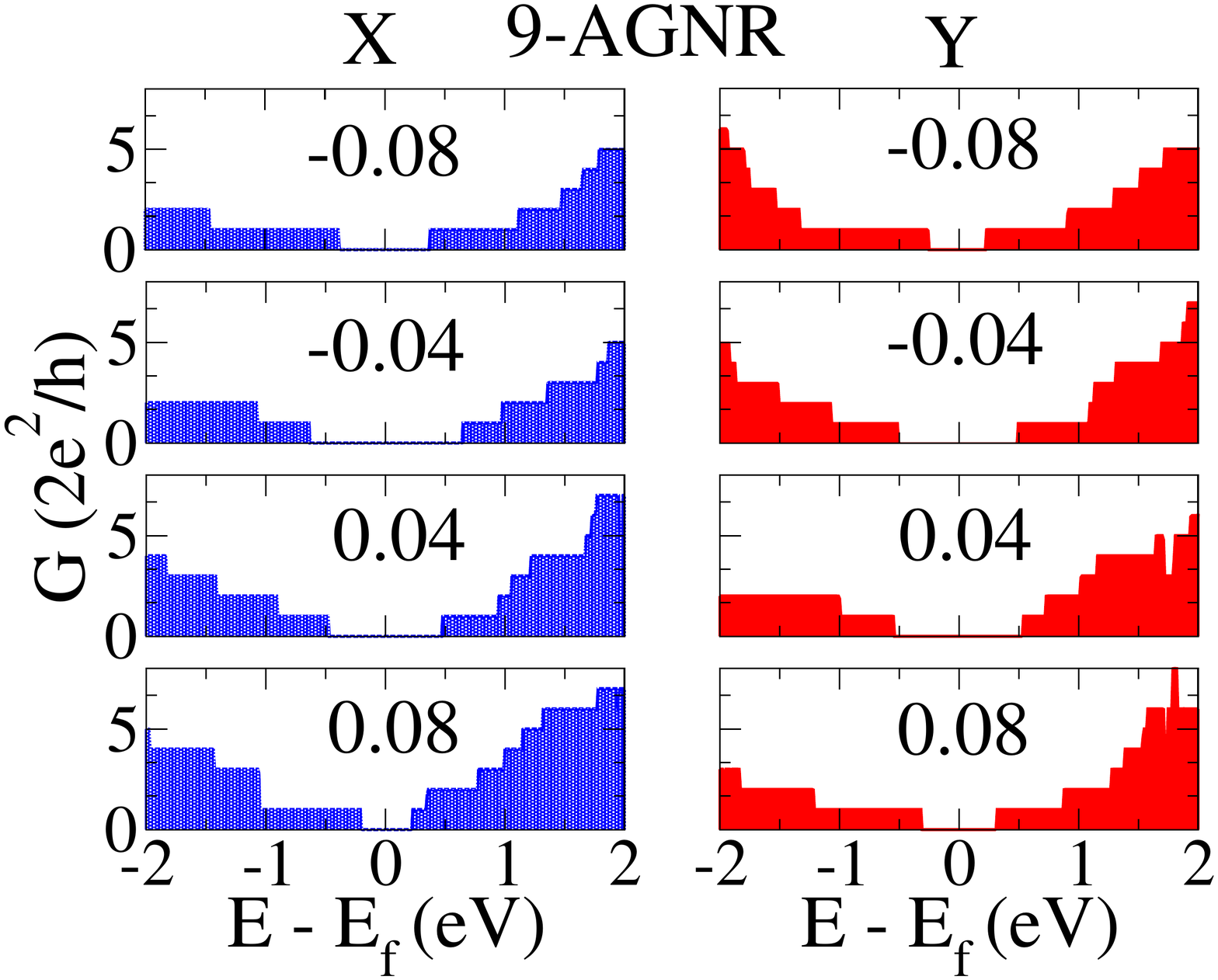}}
  \subfigure[ : The Landauer quantum conductance variation in the $x$ and $y$ direction.]{\label{10qcAGNR}\includegraphics[angle=0,scale=0.28]{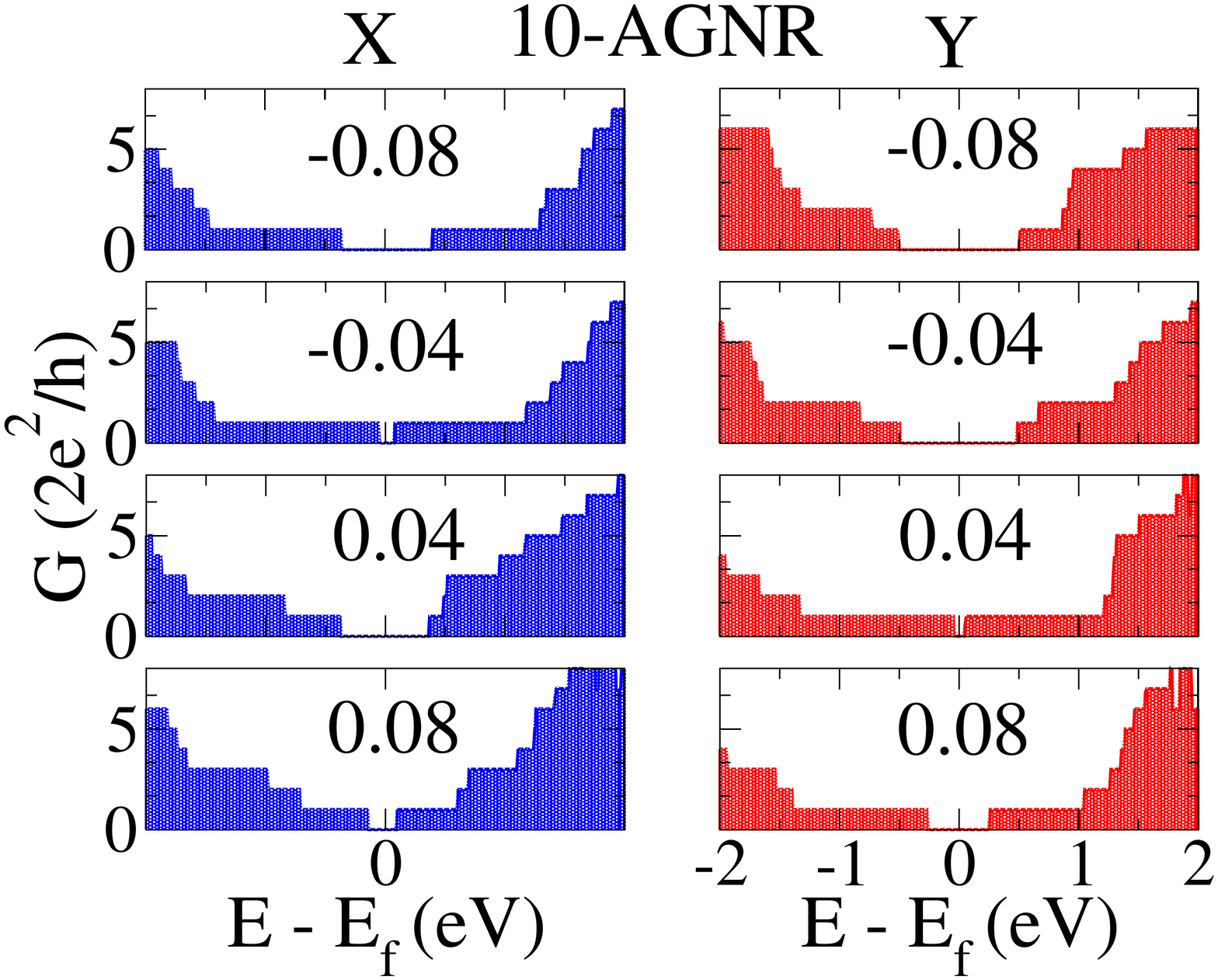}}
  \subfigure[ : The Landauer quantum conductance variation in the $x$ and $y$ direction.]{\label{11qcAGNR}\includegraphics[angle=0,scale=0.28]{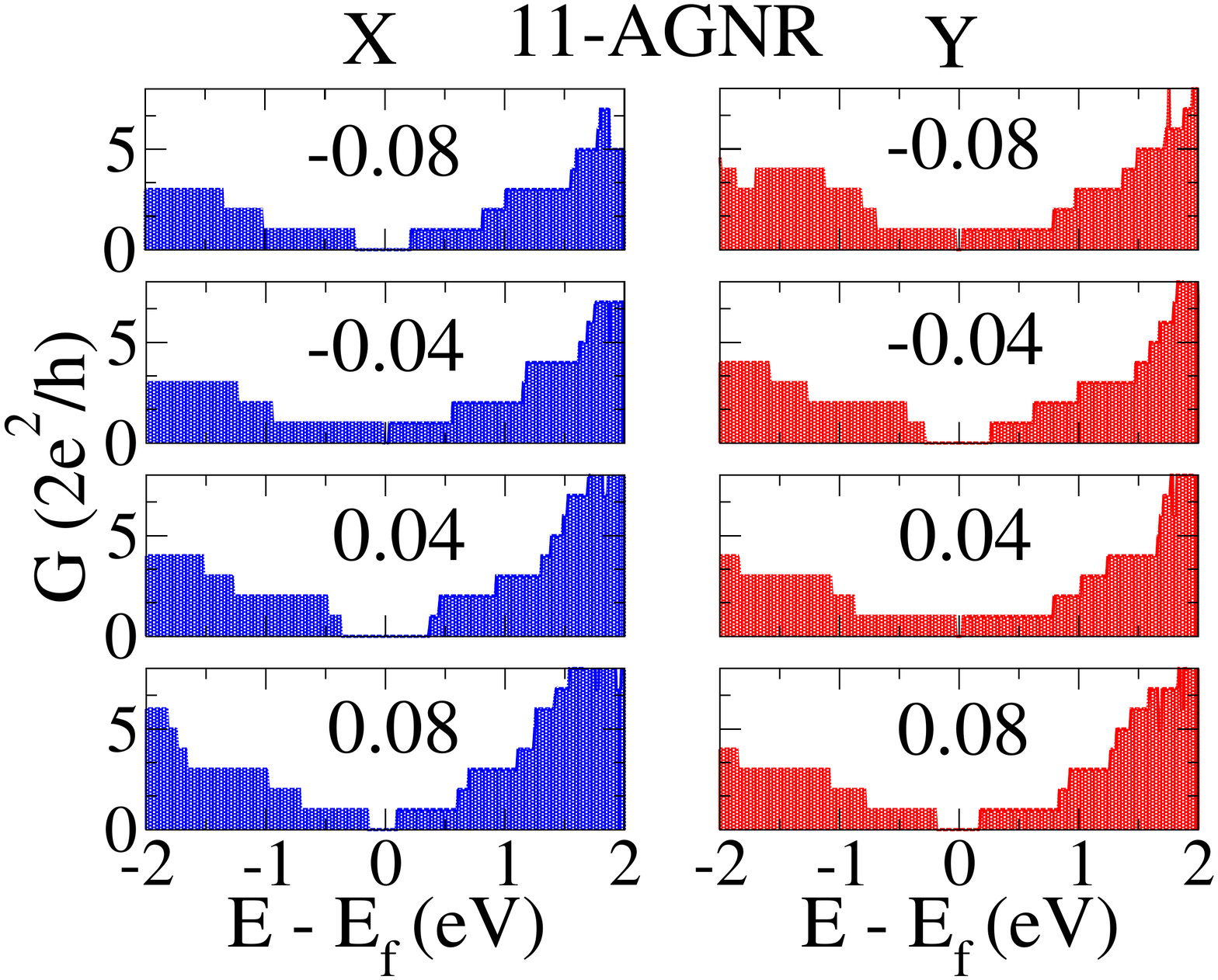}}
\end{center}
\caption{\label{conductance0} Energy gap of an AGNR under various
strains for $w$ = 3$n$, 3$n$+1 and 3$n$+2.}
\end{figure*}

\begin{figure*}[htp]
\begin{center}
  \subfigure[ : W = 9]{\label{9conduc}\includegraphics[angle=0,scale=0.17]{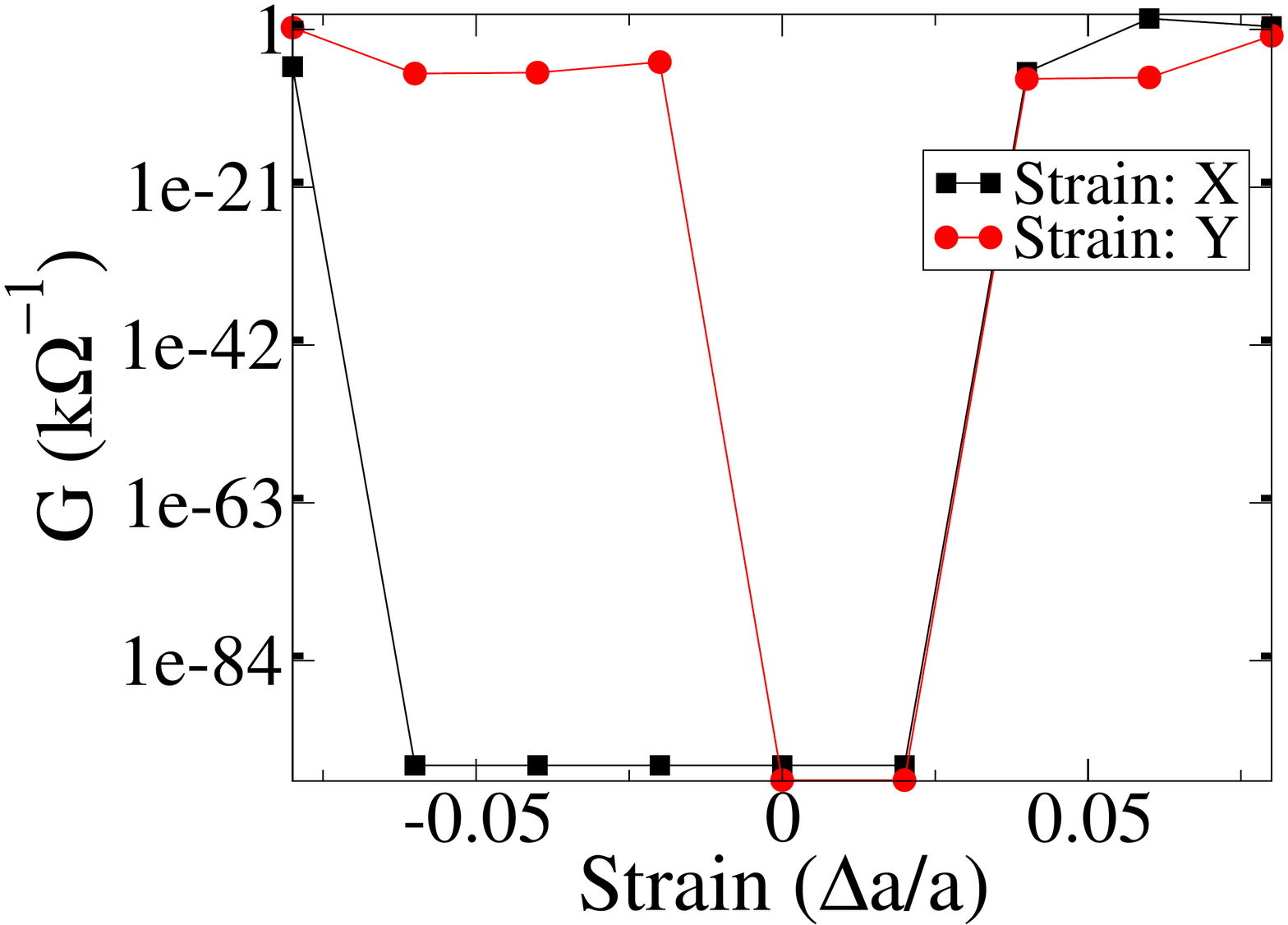}}
  \subfigure[ : W = 10]{\label{10conduc}\includegraphics[angle=0,scale=0.17]{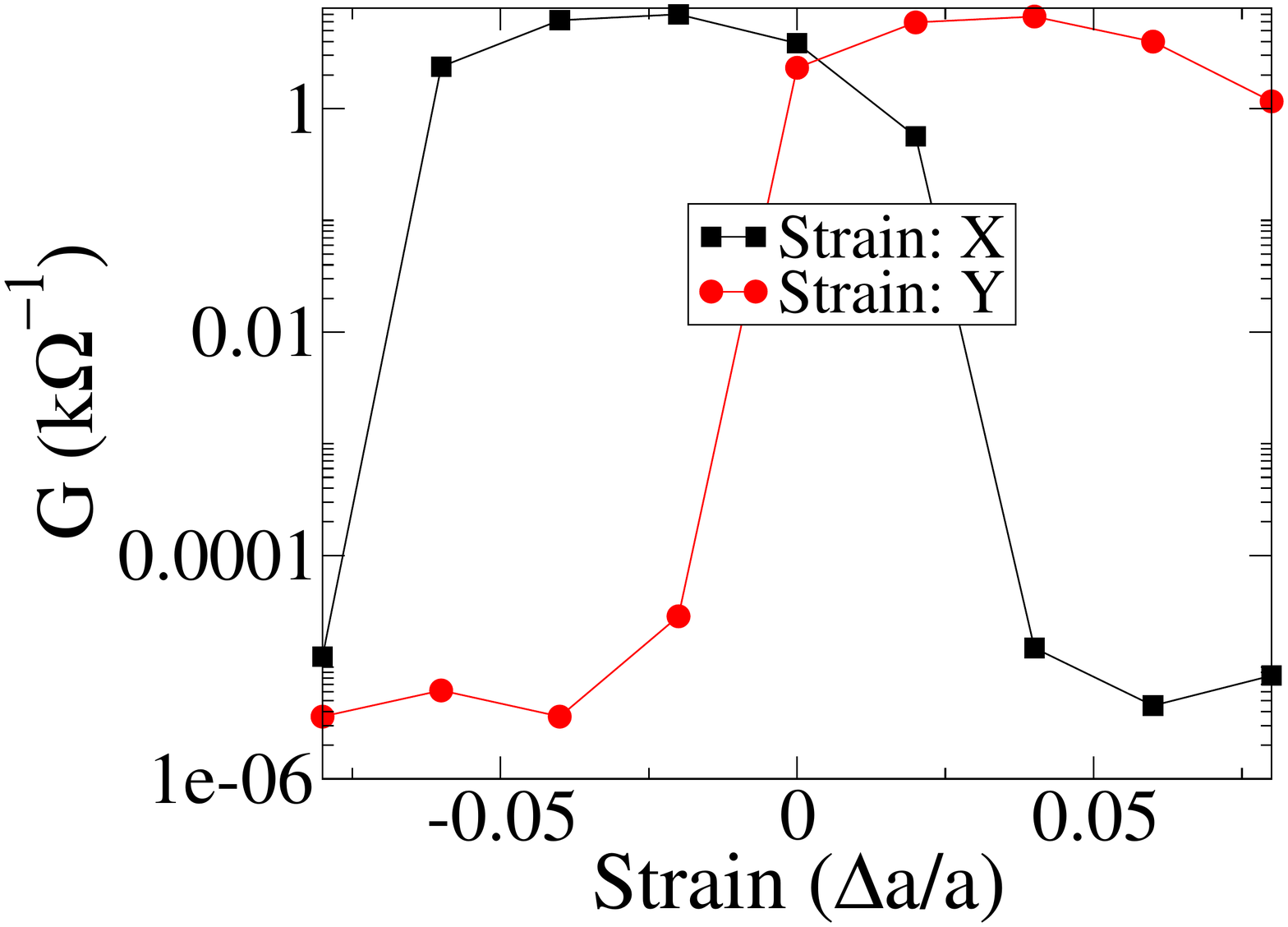}}
  \subfigure[ : W = 11]{\label{11conduc}\includegraphics[angle=0,scale=0.17]{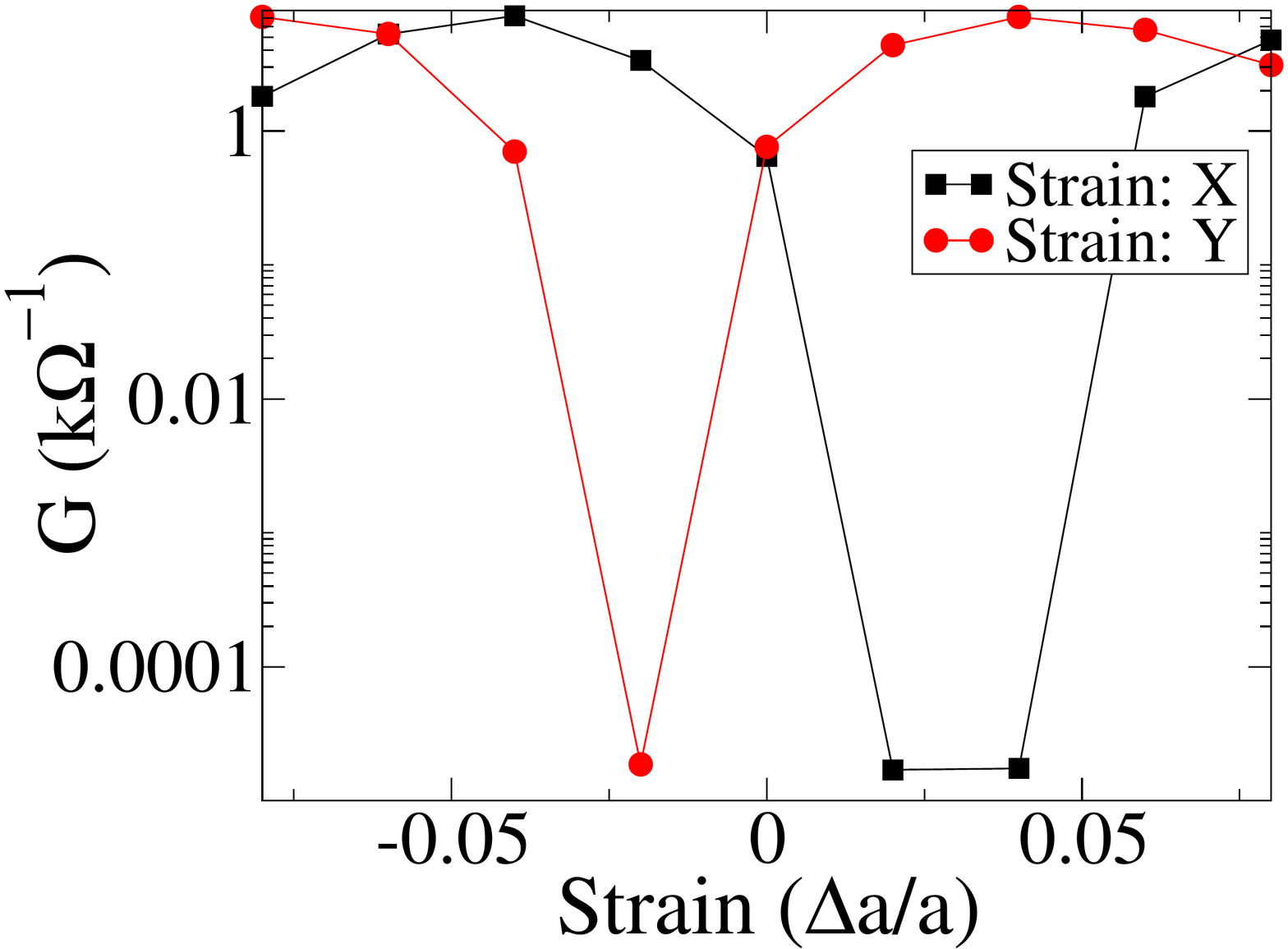}}
\end{center}
\caption{\label{conductancet} Conductance of an AGNR under various
strains for $w$ = 3$n$, 3$n$+1 and 3$n$+2 at room temperature.}
\end{figure*}

Fig.\ref{conductance0} shows the quantum conductance of the 9, 10
and 11-AGNR under various induced $\varepsilon_y$ and
$\varepsilon_x$ at T = 0 K. Only electrons at $E_f$ play a role in
the quantum conductance. The 9 and 10-AGNR, at every induced strain, have
zero conductance but for the 11-AGNR there are two special strains
($\varepsilon_x$ = -0.04 and $\varepsilon_y$ = 0.04) wherein the
ribbon conductance is nonzero due to a narrow gap. At finite
temperature, every subbands near the $E_f$ can contribute to the
quantum conductance. Current for a finite bias voltage ($V$) is
given by \cite{current}:
\begin{equation}
\label{Eq:1} I = \int_{0}^{\infty}\frac{d\varepsilon}{e}
[f(\varepsilon+\mu)-f(\varepsilon)]G(\varepsilon)\simeq\int_{0}^{\infty}\frac{d\varepsilon}{e}
[\mu\frac{\partial f}{\partial
\varepsilon}]G(\varepsilon)=V\int_{0}^{\infty}d\varepsilon
\frac{\partial f}{\partial \varepsilon}G(\varepsilon)
\end{equation}
where $f$ is the Fermi-Dirac distribution function and $\mu$ is the
chemical potential. At a finite temperature $\mu$ $\frac{\partial
f}{\partial \varepsilon}$ has a Gaussian peak that at T = 0 K
becomes a Dirac delta function. We estimate that at room temperature
the current can be approximated by integrating over $\varepsilon
\in[-0.3, 0.3]$. Fig. \ref{conductancet} shows the conductance
($G=\frac{I}{V}$) of an AGNR family versus strain at room
temperature. Because of relatively high band gap of the 9-AGNR, it
has a relatively low conductance at room temperature for a wide range
of strains (Fig \ref{9conduc}). However, the 10 and 11-AGNR have
transition between two conductance levels for the compressive and
the tensile strains. Fig. \ref{10conduc} shows the conductance of a
10-AGNR that was computed from Eq. \ref{Eq:1}. For $\varepsilon_x>0$
and $\varepsilon_y<0$, the AGNR is an insulator while for other
strain values it is a conductor. As shown in Fig. \ref{11conduc} the
conductance of the 11-AGNR for strain  values of
$\varepsilon_y=-0.02$ and $\varepsilon_x=0.02$ decreases due to the
increase of the band gap.


\begin{figure}
\center
\includegraphics[angle=0,scale=0.25]{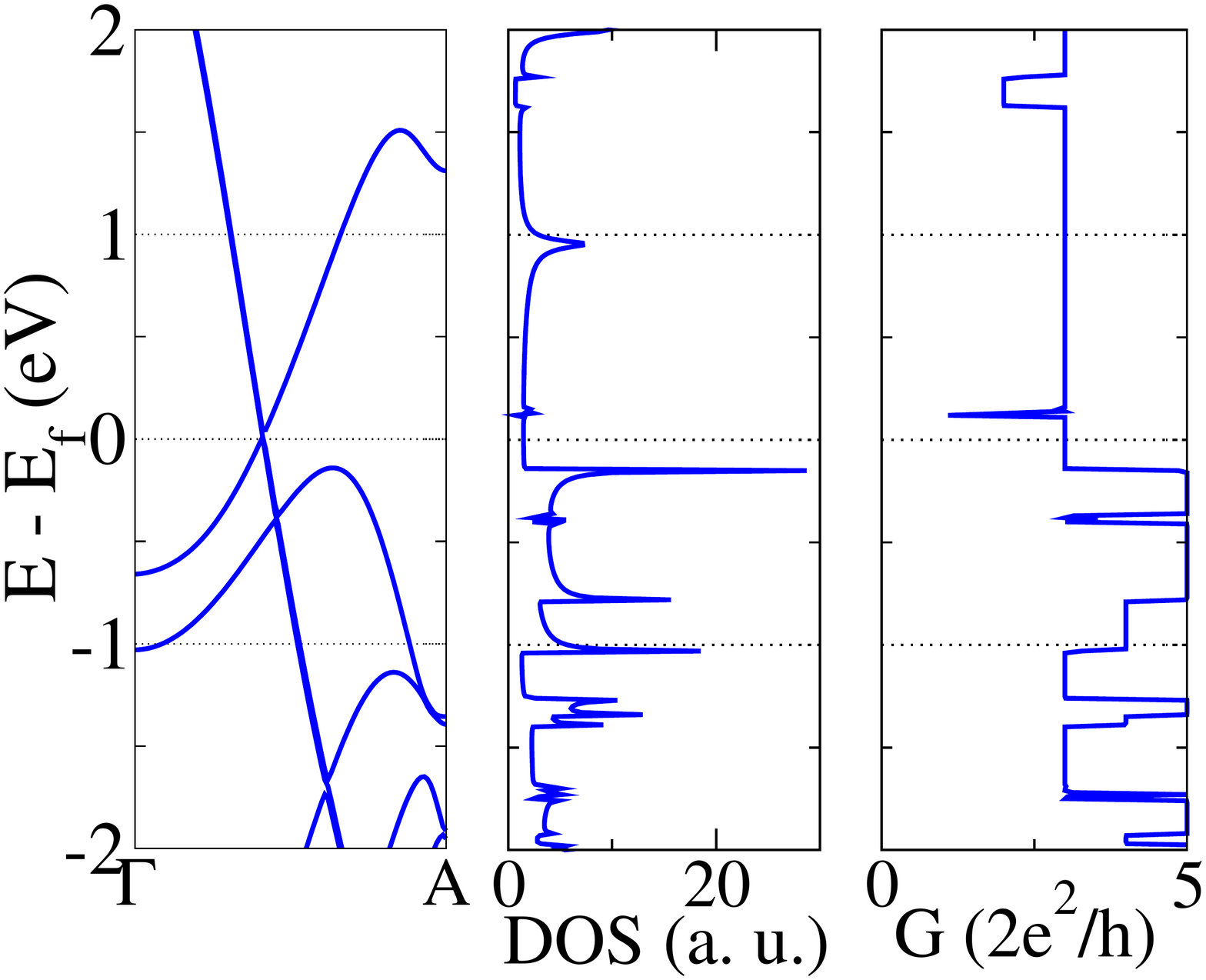}
\caption{\label{zigrelax} The band structures, the density of states
and the quantum conductance of an ZGNR at zero strain.}
\end{figure}
\begin{figure*}[htp]
\begin{center}
  \subfigure[ : Strain in $x$ direction]{\label{zigxband}\includegraphics[angle=0,scale=0.25]{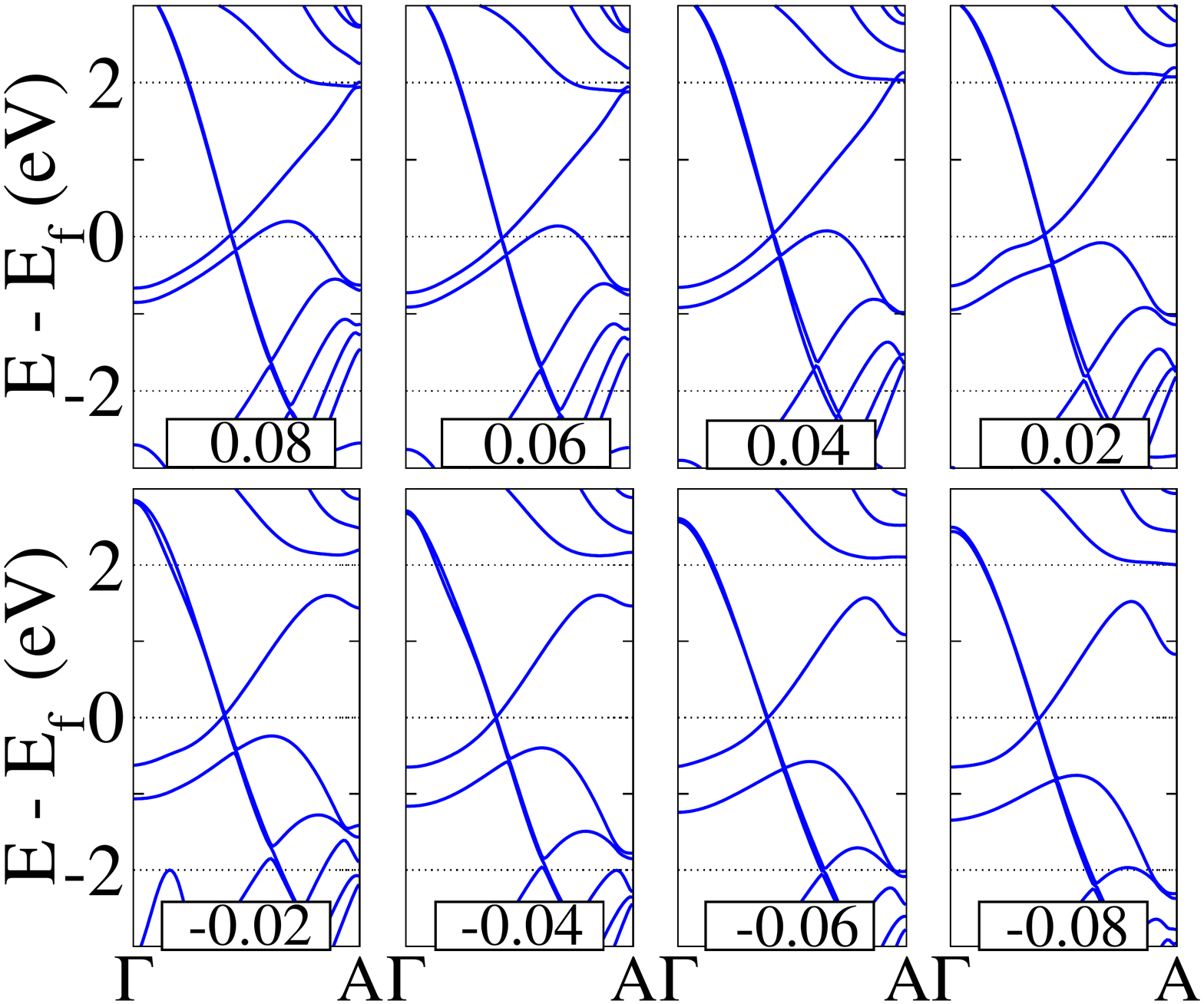}}
  \subfigure[ : Strain in $y$ direction]{\label{zigzband}\includegraphics[angle=0,scale=0.25]{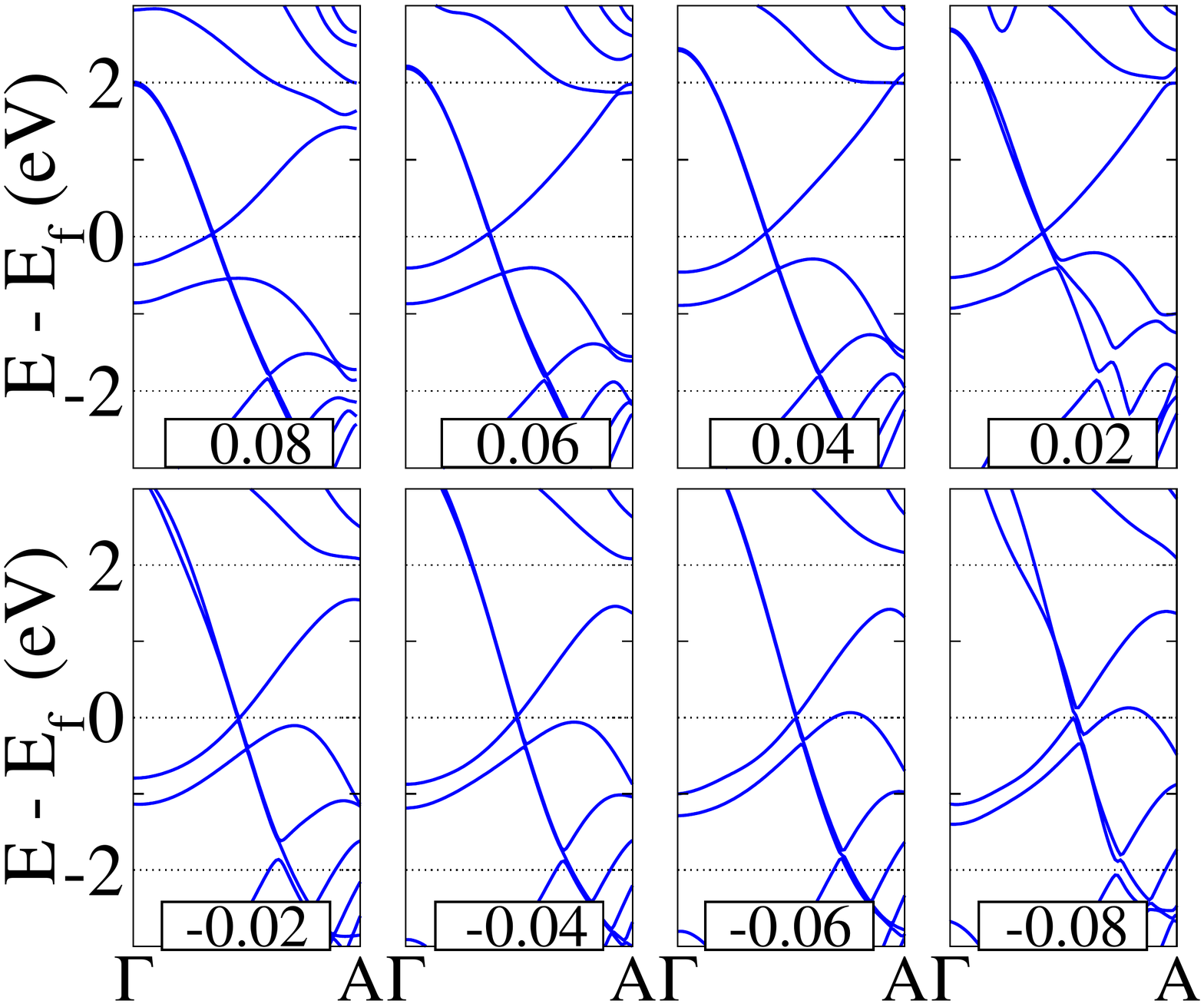}}
  \subfigure[ :  The Landauer quantum conductance variation by $x$ direction strain.]{\label{zigtransx}\includegraphics[angle=0,scale=0.25]{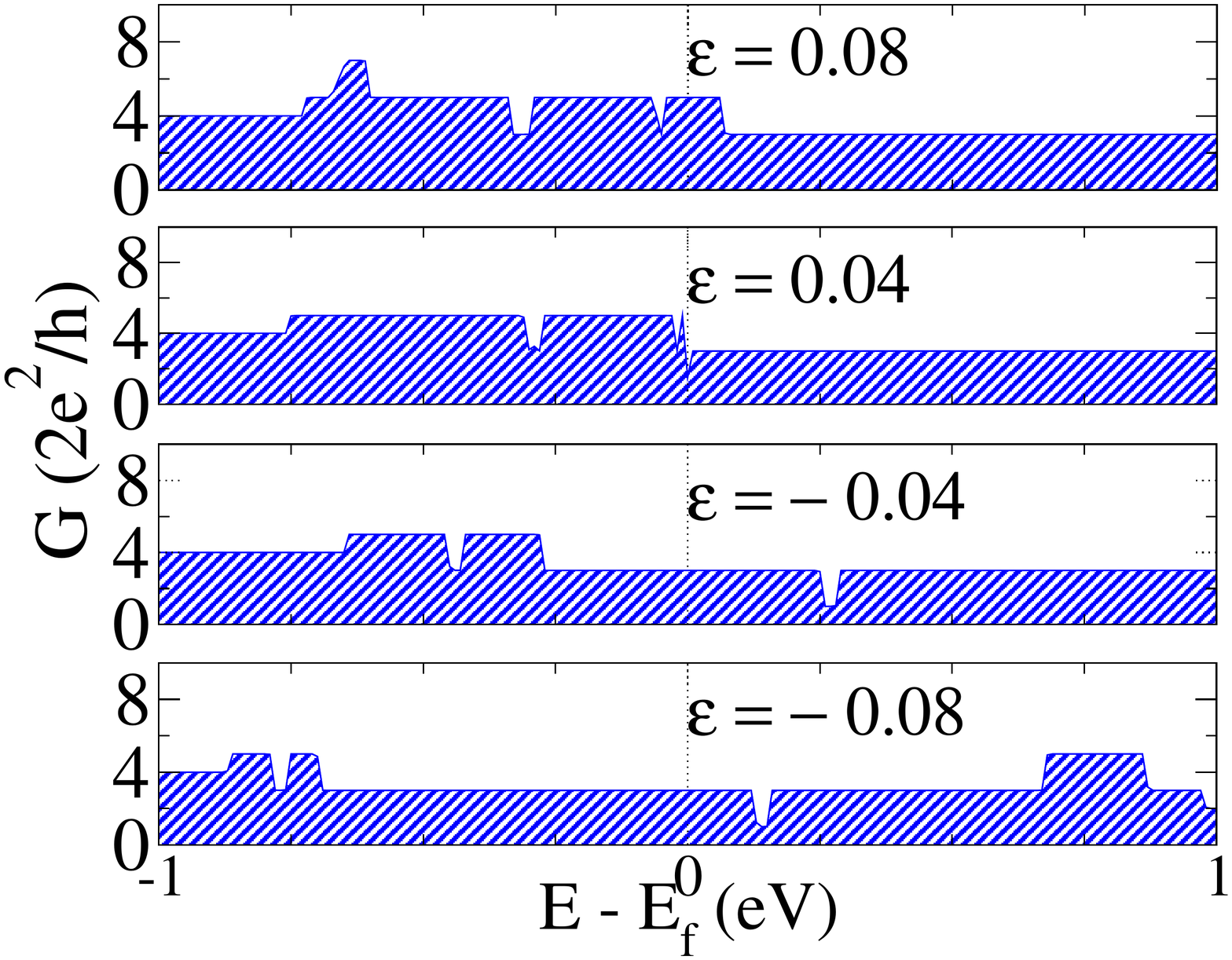}}
  \subfigure[ : The Landauer quantum conductance variation by $y$ direction strain.]{\label{zigtransz}\includegraphics[angle=0,scale=0.25]{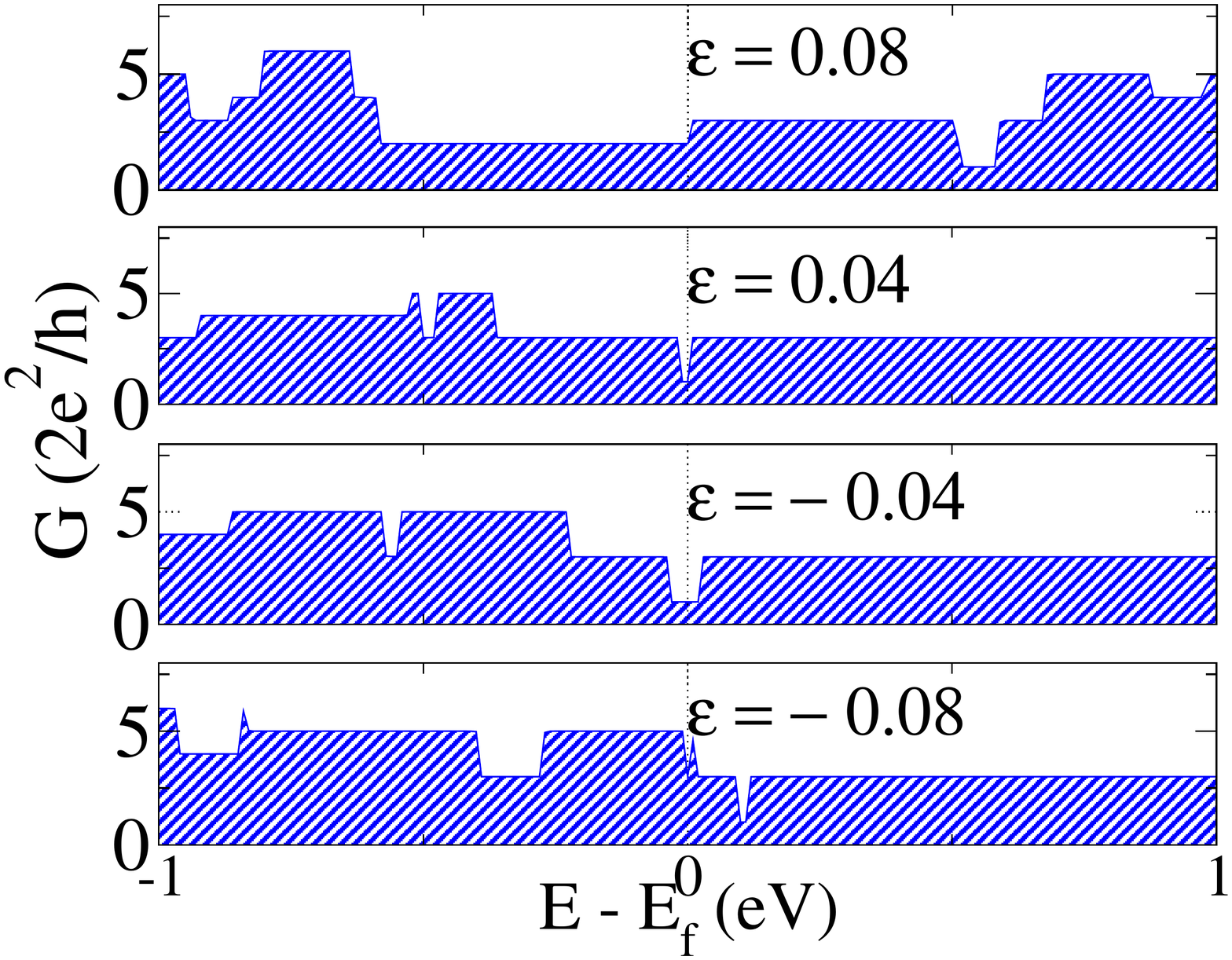}}
\end{center}
\caption{\label{bandzig} The band structures and the quantum
conductance of an ZGNR under various strains.}
\end{figure*}

We now present the electronic structure of an ZGNR at zero strain.
Fig. \ref{zigrelax} presents the band structures, the density of
states and the quantum conductance of an ZGNR. The band structure
near the $E_f$ has a linear dispersion and the group velocity near
this level is about $2.1 \times 10^6$ $ms^{-1}$. To calculate the
conductance and the density of states we construct the wave
functions in the energy window with $E \in [-6,6]$ eV around the
$E_f$ that is taken to be the reference zero. Quantum transport of
the ZGNR near the valence band edge is about $6e^2/h$.

Fig. \ref{bandzig} shows the band structure of a ZGNR under various
uniaxial strains. The induced $\varepsilon_x$ cause an increase in
the group velocity of the Dirac fermions from $2.0 \times 10^6$
$ms^{-1}$ to maximum $2.3 \times 10^6$ $ms^{-1}$ for the strain from
-0.08\% to 0.08\% (Fig. \ref{zigxband}), but the induced
$\varepsilon_y$ cause an decrease in the group velocity from $2.7
\times 10^6$ $ms^{-1}$ to minimum $2.0 \times 10^6$ $ms^{-1}$ for
the same strain range (Fig. \ref{zigzband}). The variation in the
band structures influence the electrical conductance of the GNR.
Fig. \ref{zigtransx} and \ref{zigtransz} show the quantum
conductance of the ZGNR under various $\varepsilon_x$ and
$\varepsilon_y$. By comparing these results with those related to
the relaxed ZGNR (Fig. \ref{zigrelax}) we find that the strain can
increase the conductance of the ZGNR up to two times. The
compressive $\varepsilon_x$ shifts some bands to below the $E_f$ but
doesn't change the quantum conduction near the $E_f$, while the
tensile $\varepsilon_x$ enhances it to $10e^2/h$. In the $y$
direction, the ZGNR is very sensitive to the strain and the tensile
$\varepsilon_y$ causes an increase in the conductance, while the
compressive $\varepsilon_y$ decreases it at first but increases
later.

\section{Conclusions}
\label{conclusions} Density functional studies of strain effects on
the electrical transport properties of the graphene nanoribbons are
presented. By applying a uniaxial tensile strain in the $x$ and $y$
directions, the electronic properties of the graphene nanoribbons
were studied. Using the Wannier functions, the band structure and
density of states were calculated for different strains from $-8\%$ to
$8\%$. It is observed that as the strain increases, depending on
AGNR family type, the electrical conductivity changes from an
insulator to a conductor, and this is accompanied by the variation in
the electron and hole effective masses. Compressive $\varepsilon_x$ in the
ZGNR shifts some bands to below the $E_f$ and the quantum conductance
does not change but the tensile $\varepsilon_x$ causes an increase
in the quantum conductance to $10e^2/h$ near the $E_f$. In the
transverse direction, the ZGNR is very sensitive to the strain and the
tensile $\varepsilon_y$ causes an increase in the conductance while
the compressive $\varepsilon_y$ decreases it at first but increases
later.

{\bf Acknowledgment:}

H. R-T. acknowledges the support from Iran National Science
Foundation. We thank Dr. M. Farjam for useful discussions.
\bibliographystyle{apsrev}
\bibliography{bibliography}
\end{document}